# Flow Rate Independent Multiscale Liquid Biopsy for Precision Oncology

Jie Wang[1], Robert Dallmann[2], Renquan Lu[3], Jing Yan*[4], and Jérôme Charmet*[2,5,6]

1 Institute for Advanced Materials, School of Material Science and Engineering, Jiangsu University, Zhenjiang 212013, China,
2 Division of Biomedical Sciences, Warwick Medical School, University of Warwick, Coventry CV4 7AL, United Kingdom,
3 Department of Clinical Laboratory, Fudan University Shanghai Cancer Center, Shanghai, 200032, China,
4 Holosensor Medical Technology Ltd., Suzhou, 215000, China,
5 WMG University of Warwick, Coventry CV4 7AL, United Kingdom,
6 School of Engineering – HE-Arc Ingénierie, HES-SO University of Applied Sciences Western Switzerland, 2000 Neuchâtel, Switzerland.

* corresponding authors: yj@holosmed.com (+86 400 1816 488) and jerome.charmet@he-arc.ch (+41 32 930 2629)

## Abstract

Immunoaffinity-based liquid biopsies of circulating tumor cells (CTCs) hold great promise for cancer management, but typically suffer from low throughput, relative complexity and post-processing limitations. Here we address these issues simultaneously by decoupling and independently optimizing the nano-, micro- and macro-scales of an enrichment device that is simple to fabricate and operate. Unlike other affinity-based devices, our scalable mesh approach enables optimum capture conditions at any flow rate, as demonstrated with constant capture efficiencies, above 75% between 50–200 µL min$^{-1}$. The device achieved 96% sensitivity and 100% specificity when used to detect CTCs in the blood of 79 cancer patients and 20 healthy controls. We demonstrate its post processing capacity with the identification of potential responders to immune checkpoint inhibition therapy and the detection of HER2 positive breast cancer. The results compare well with other assays, including clinical standards. This suggests that our approach, which overcomes major limitations associated with affinity-based liquid biopsies, could help improve cancer management.

## Introduction

Liquid biopsies have the potential to transform cancer management through non-invasive, real-time feedback on patient conditions [1]–[4]. Circulating tumor cells (CTCs) that are released from primary and/or distant tumors into the bloodstream [5], are seen as a particularly useful source of information to improve clinical outcomes (patient prognosis, real-time responses to therapeutic interventions, monitoring of tumor recurrence), guide drug discovery, and advance our understanding of cancer progression, metastatic cascade, and minimal residual diseases [6]–[9]. However, the capture of such cells from blood is technically challenging due to their low abundance, typically 1-10 CTCs per mL [10]–[12]. This constraint imposes the processing of a large sample volume, usually between 4-10 mL, to retrieve enough cells. Consequently, the ideal device must combine high capture efficiency and high throughput.

Amongst the many CTC enrichment strategies, those relying on size differences to discriminate cancer cells from healthy blood cells have recently gained in popularity [3]. This is because these approaches are capable of very high throughput using devices that are relatively simple to fabricate and operate. However, it has become apparent that the CTCs captured by such devices may fail to recapitulate their native biological complexity and heterogeneity. For example, such devices may miss small-sized CTCs [13] that are correlated with aggressive metastatic progression in patients [14]. Even though several multi-step solutions have been developed recently to address these issues [15]–[18], they are more complex than their counterparts based on physical capture alone. Therefore, it would be ideal to develop an affinity-based solution that harbors the same simplicity and throughput as their cells-size based counterparts.

In affinity-based devices, the capture of cells typically relies on the interaction between the cell's surface markers and complementary antibodies tethered to the channel walls. The current issue with such devices is the reliance on channels or structures with dimensions on the order of the target cells size. Even though small channels dimensions enhance the interaction between the target cells and the capture elements [2], [3], [10], [19]–[21] they also limits the flow rate and hence the throughput. Indeed, most affinity-based microfluidic devices proposed to date have an upper limit of a few mL h$^{-1}$, above which the capture efficiency drops significantly (typically up to 2 mL h$^{-1}$ as reported in a number of reviews [3], [11], [22], [23]). In addition to the flow rate limitation, affinity-based liquid biopsy devices are usually complex due to their inherent small sizes and channel geometries (e.g., [23]–[25]). Finally, such devices do not always allow for easy post-processing since the cells are typically surface-bound inside the chip and not readily accessible or retrievable.

Our study, which revisits widely accepted yet misleading common knowledge in microfluidics, allowed us to propose a novel strategy that addresses these issues. We conceptualised a simple yet widely applicable solution that relies on a scalable macroscale mesh with nano-functionalised micropores (Fig 1a-d). Our solution addresses the 1) flow rate dependence, 2) complexity and 3) post-processing limitations simultaneously. First, we demonstrate an optimized capture efficiency above 75%, independent of the flow rate, up to 200 µL min$^{-1}$ (or 12 mL h$^{-1}$), which is approximately an order of magnitude higher than most commonly reported values for surface-based capture in conventional microfluidic devices [11], [22], [24]–[26]. Second, the production of our device does not rely on complex microfabrication processes and its operation only requires an external pumping system to process manually loaded buffers and samples (Fig. 1e-i). Third, it allows for easy, off-chip functionalization before assembly, and importantly, allows for simple post-processing of the captured CTCs. Figure S1 shows the detailed process steps.

We validate the device clinically by isolating CTCs in 4 mL blood samples from 79 cancer patients. The sensitivity and specificity of our device using conventional staining protocols to identify the CTCs were of 96% and 100%, respectively against a group of healthy controls (n=20). We also demonstrate its post processing capacity with the identification of potential responders to immune checkpoint inhibition therapy and the detection of HER2 positive breast cancer. The results compare well with other assays, including clinical standards, and show the potential applicability of our simple multiscale, flow rate independent liquid biopsy strategy for cancer management.

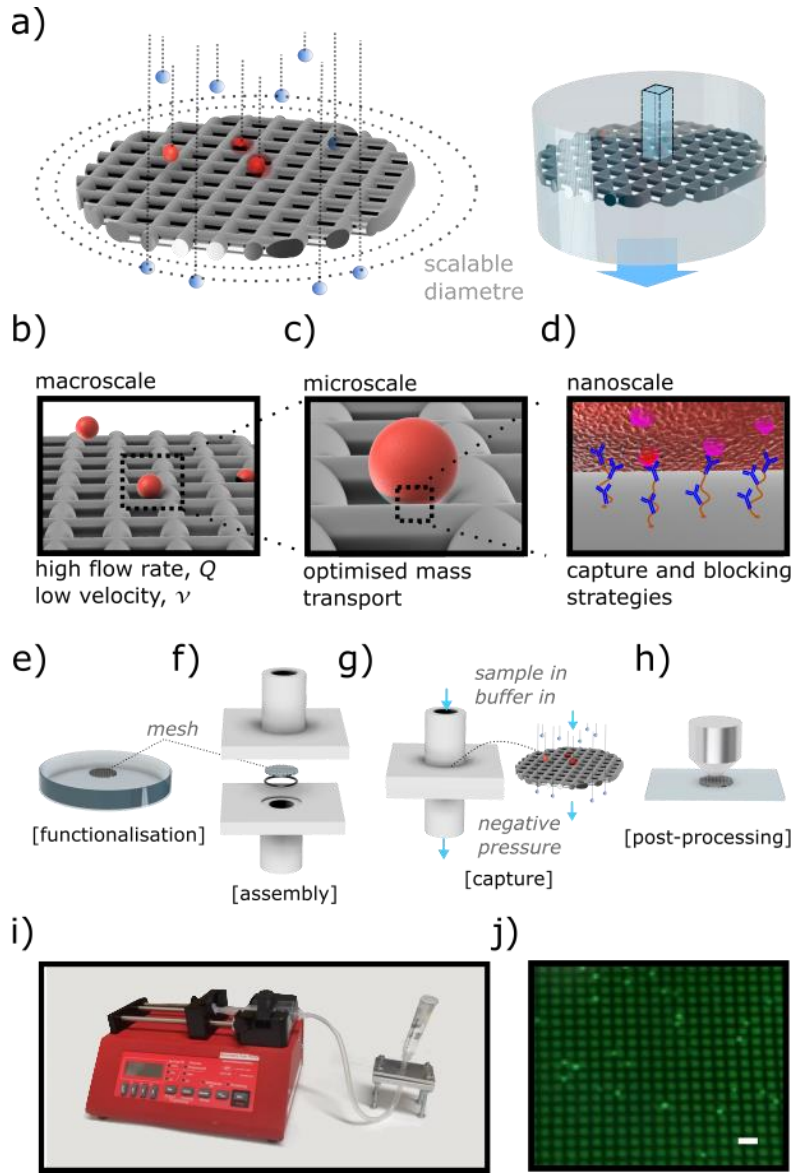

***Figure 1****. **Architecture of the device***. *The core of the device is a scalable micromesh – not to scale (**a**) that enables the decoupling and independent optimization of biosensing-relevant length scales. The resulting device offers a macroscale channel to maintain low fluid velocity while running samples at high flow rate (**b**), a microscale mesh whose dimension enhances interactions with the target cells (**c**) and a nano-functionalized coating that enables high capture efficiency and low non-specific interaction (**d**). The main operations including functionalization (**e**), assembly (**f**), capture (**g**) and post-processing (**h**) are made easier by the fact that the mesh can be easily removed from its holder, allowing access for pre- (**e**) and post-processing (**h**) as shown in (**j**) that displays GFP expressing MCF-7 cells observed after capture, scale bar 40 μm. The assay can be performed using widely available laboratory equipment (**i**).*

**Results**

**Decoupling length scales for flow rate independent capture and device optimization**

To better understand the flow rate limitation that seems to affect affinity-based solutions, we concentrate on the Peclet number (*Pe*), a dimensionless metric used to optimize biosensors [27]–[29]. The Peclet number measures the ratio of the convection rate over the diffusion rate. Applied to the context of rare events (such as CTCs in blood), a small Peclet number (<<1) is ideal as it promotes interaction between the target cell and the functionalized surface. In other words, it is a condition that maximizes cell capture. In most microfluidics publications, the Peclet number is written as $Pe=Q/DL$ (e.g., [27], [28], [30]), $Q$ being the flow rate, $D$ the diffusion coefficient, and $L$ the characteristic length scale. This notation emphasizes the flow rate limitation. Indeed, the equation suggests that flow rate cannot be increased without decreasing the Peclet number (and hence the capture efficiency). However, the Peclet number can also be written as $Pe=vL/D$, which implies a fluid velocity (*v*) limitation. This notation allows us to conceptualize a situation with a low Peclet number and high flow rate, provided the latter can be decoupled from the fluid velocity.

The fact that the Peclet's number flow rate limiting notation is so widely accepted is because flow rate and velocity are coupled via the channel's cross section in conventional channels (see $v=Q/A$ and Supp. Mat. for details). However, it suffices to introduce structures that have two inherent length scales to decouple the two parameters. Here we use meshes with microscale pore size and macroscale diameter mounted in a channel of matching size. In this configuration, the velocity can be kept low (to maintain $Pe \gg 1$ and hence optimal capture) for any flow rate. Indeed, for a given flow rate, the fluid velocity can be reduced by increasing the cross section of the mesh $A_s$, as shown in $v=Q/A_s$. We achieve this by cutting the mesh to the right diameter, which does not compromise the micron-size length scale (pore size) necessary for optimal mass transfer or the nanoscale critical for functionalization strategies.

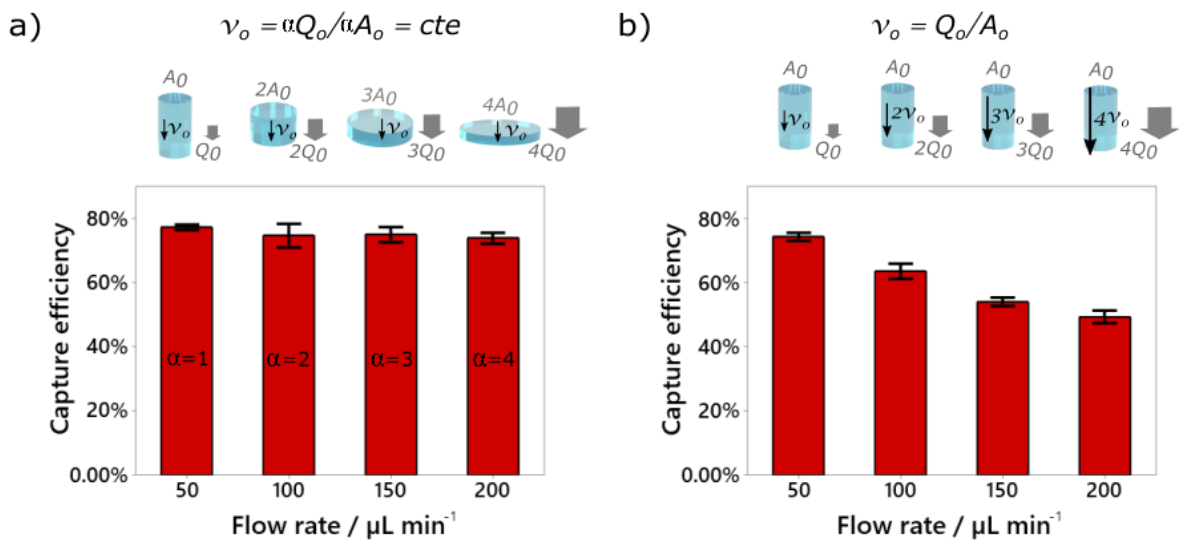

**Figure 2. Flow rate independence**. *Panel (**a**) shows that capture efficiency can be kept constant (as confirmed by one-way ANOVA [F=0.37, p=0.05] and post-hoc Tukey's test (Fig. S2)) as long as the velocity is kept constant. This is achieved by scaling the surface area by the same factor as the flow rate as given by $v = \alpha Q/\alpha A$. In comparison, one observes a significant capture efficiency decrease (one-way ANOVA [F=38.04, p=0.05]), when the diameter of the mesh is kept constant (**b**). The measurements were performed using MCF-7 cell lines spiked in buffer solutions. The bars represent the standard error.*

Taking the concept even further, we show how to achieve a flow rate independent capture. Indeed, an arbitrary velocity $v_a$ can be kept constant provided the flow rate and the diameter are scaled by the same factor α as shown in $v_a = αQ_a/αA$. This is demonstrated experimentally by measuring the device's capture efficiency at different flow rates (Fig. 2). The capture efficiency is given by the ratio of the captured to the introduced target cells. We used GFP expressing MCF-7 breast cancer cell line (as shown after capture in Fig 1j) and the functionalization procedure described below on a 15 x 20 µm pore size mesh. First, we defined the diameter-dependent optimal flow rate $Q_o$. It is the flow rate yielding the maximum capture efficiency before drop-off, using a mesh with a fixed diameter. A constant capture efficiency of 75% is observed until $Q_o = 50$ µL min⁻¹ before decreasing significantly as shown in Figure 2b (and Fig. S3 in Supplementary Materials), for a mesh of 6 mm diameter. This behavior is consistent with other affinity-based liquid biopsies [11], [22], [24] and with our simulations (Fig. 3). The optimal velocity of our system, $v_o = Q_o/A_s$ (with $Q_o = 50$ µL min⁻¹ and $A_s \cong 88.3\ mm^2\ (\emptyset = 6\ mm)$) is thus $v_o = 2.95$ x 10⁻⁵ m s⁻¹. This velocity can be kept constant for any flow rate provided $Q_o$ and $A_s$ are multiplied by the same factor, α. Figure 2a shows no significant difference in capture efficiency for α = 1 - 4, i.e., for flow rates ranging from 50 to 200 µL min⁻¹, as determined by one-way ANOVA [F=0.37, p=0.05]. A post-hoc Tukey's test shows that there is no statistical difference between any of the flow rates (Fig. S2). This result is in stark contrast with Figure 2b that shows a strong dependency on the flow rate, as determined by one-way ANOVA [F=38.04, p=0.05], when the diameter of the mesh is kept constant. Graphs with flow rates down to 20 µL min⁻¹ (Fig. S3) and details of the post-hoc Tukey's tests (Fig. S2, S4) are provided in Supplementary Materials. More practically, this approach can be used to find an optimum capture efficiency given a target flow rate, $Q_t$. Indeed, a simple rule of three suffices to define the optimal mesh cross section, $A_{st}$, as defined by $A_{st} = Q_t/v_0 = A_s\ Q_t/Q_0$ (the initial surface area times the ratio of the target to the initial flow rate).

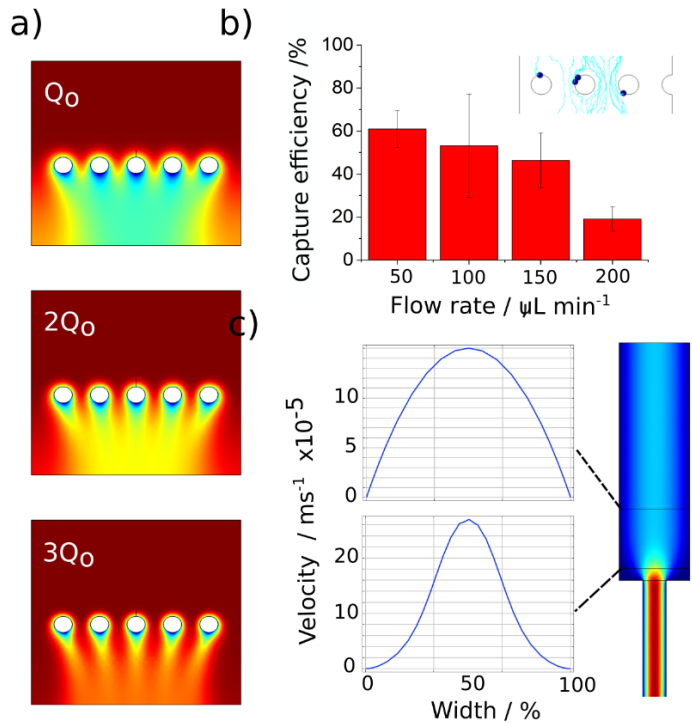

***Figure 3. Comparative (semi-quantitative) multiphysics simulations***. *Panel (**a**) shows the effect of the flow rate on the capture of diluted species (arbitrary concentration $C_0$ indicated in red) on the mesh. Here a cross section is represented. A depletion zone (blue) appears for a flow rate $Q_0$ and decreases progressively as the flow rates are increased (2$Q_0$ and 3 $Q_0$). The same was observed for discrete events,*

*using particle tracing in (**b**). In this case, we performed individual measurements (at least triplicates) and reported them in a graph to show the flow rate dependency on the capture efficiency. Panel (**c**) shows the effect of restrictions in the vicinity of the mesh (cross section). If the mesh is too close to a diameter restriction, the velocity increases locally, thereby negatively impacting the capture efficiency.*

Comparative (semi-quantitative) multiphysics simulations (COMSOL Multiphysics 5.5) were performed to optimize the device (Fig. 3). The effect of the flow rate on the capture efficiency is shown in figures 3a,b, which represent the cross section of a mesh in a channel. Figure 3a displays the reaction of diluted species on the mesh. A diluted solution of particles (arbitrary concentration $C_0$) is represented in red and introduced at the top of the channel at arbitrary flow rates $Q_0$, $2Q_0$ and $3Q_0$. The background solution is represented in blue ($C_0 = 0$). The blue traces therefore represent the solution depleted from the particles captured on the mesh. The simulations show that the number of particles captured decreases with increasing flow rate (i.e., less material is captured by the mesh) for a given channel diameter, which confirms our experimental results (Fig 2b). The particle tracing module was used to evaluate the flow rate dependence of the capture efficiency for discrete events, to represent individual cells. Figure 3b shows a representative image (inset) and a graph based on a series of individual repeats and confirms the previous observations (experiments and simulations). In both cases, several mesh parameters, including wire diameter and pore sizes, were evaluated to select the meshes for experimental work. The results provided qualitative data (not shown) whose trend aligned well with experiments. We also used the simulation to optimize the position of the mesh in the device. Figure 3c highlights the effect of channel diameter restrictions, which can locally increase the velocity across a mesh if it is positioned in its proximity. Since an increased velocity reduces the capture efficiency, it is important to position the mesh sufficiently far away from any diameter restrictions.

**Nanobranched polymer and capture optimization**

We used thiol-terminated nanobranched polymers, tethered to a gold coated micromesh, functionalized with antibodies against specific cell surface receptors to confer high capture efficiency, and high specificity to our device. First, we targeted epithelial cell adhesion molecules (EpCAM) that are characteristically overexpressed in a range of epithelial cancers, but not in other blood cells [31]. EpCAM has already been used in a number of affinity-based liquid biopsy studies [11], [22], [31]–[33] and therefore provides a good standard for comparison. Nanobranched polymers can accommodate multiple anti-EpCAM antibodies alongside blocking molecules, increasing the probability of antigen-to-antibody contact and minimizing non-specific interaction [34]–[36].

The simple design of the device enables seamless removal and mounting of the microscale mesh (Fig. 1e-h), such that its functionalization can take place in optimal conditions, with minimum waste of precious material. Figure 4a represents the simplified polymer synthesis steps. Detailed descriptions of the polymer synthesis and functionalization steps are given in the Methods. Briefly, gold coated micromeshes were incubated in sulfhydryl hyaluronic acid (HA-SH) for two hours, followed by activation using EDC/NHS in MES buffer (pH=6). After 30 min, anti-EpCAM antibodies were incubated at 37 °C for 2 h to finalize the functionalization of the mesh. Antibody dosage and incubation times were systematically examined to optimize binding efficiency (Fig. S5 in Supplementary Materials). After the reaction, blocking molecules were added (1 h incubation time) to reduce non-specific binding (details provided below).

We first optimized our device using buffer solutions spiked with a known number of EpCAM expressing MCF-7 cells. The capture efficiency of the device is given by the ratio of captured cells to the total number of cells added to the solution. Before capturing, MCF-7 cells were loaded with intra-cellular live cell dye (CFSE) to simplify their counting and differentiation from background. To optimize the cell culture parameters, we compared the effect of different harvest reagents and the influence of passage number on the EpCAM receptor's integrity, by performing flow cytometry and immunostaining (Fig. S6). Having optimized these parameters, we evaluated the capture efficiency.

A functionalized 6 mm diameter mesh was mounted into the holder with a fluidic channel of matching ($\varphi = 6$ mm) dimension. A syringe pump was then connected to the holder via medical grade tubing. After priming the device with buffer to the top of the mesh, medium spiked with appropriate cells was added to the open reservoir. The solution was then withdrawn through the mesh at a flow rate of 50 µL min$^{-1}$, which corresponds to the optimal flow rate for a 6 mm diameter mesh as explained above. Next, fresh medium was added to the reservoir to wash off non-specifically bound cells and ensure that only the cells captured by affinity binding stay on the mesh. Finally, the mesh was removed and observed under a microscope to count the captured cells. Figure 4b shows cells captured on the mesh with varying initial cell concentrations. The simplified process steps are shown in Fig. 1e-h (a more detailed version is presented in Fig. S1).

Capture efficiency of 58% ($\sigma^2 = 6.3\%$) is obtained with a $20 \times 20$ µm pore size micromesh (Fig. 4c). In contrast, antibodies tethered directly to the gold coated mesh via Traut reagent result in capture efficiencies below 20% (Fig. S7). These result are in line with previously reported studies that noted improved capture efficiency when using nanobranched polymers [36]. Figure 4c also shows that the capture efficiency increased with decreasing pore sizes (at 0.05 level, the population means are significantly different, using one-way ANOVA, F=15.12), even though the $10 \times 18$ µm and $15 \times 20$ µm pore size meshes cannot be considered statistically different using post-hoc Tukey's test (Fig. S8). This increased efficiency is attributed to the higher probability of cells interaction with the functionalized surface. Micromeshes with $10 \times 18$ µm pore sizes result in capture efficiency up to 81.6% ($\sigma^2 = 1.4\%$), and correspond to typical values reported for a range of affinity-based microfluidic approaches [37], [38]. Even though such pore sizes can also filter larger CTC clusters, they are generally too large for capturing single CTCs [39], [40]. This further confirms that the captures observed in our case are due to affinity binding. It is interesting to note that the capture efficiency does not scale linearly with the projected surface area of the mesh, i.e., the active functionalized surface as 'seen' by the cells. The reduction observed for smaller mesh pore size is attributed to higher local velocity due to the decrease in total open area compared to the optimized flow rate for larger pore size. Indeed, the total mesh open area is reduced by 19% with the $10 \times 18$ µm pore sizes compared to the $20 \times 20$ µm pore size, resulting in a velocity increase of 24% for the same flow rate. This suggests that the flow rate $v_o$ should be optimized for each pore size.

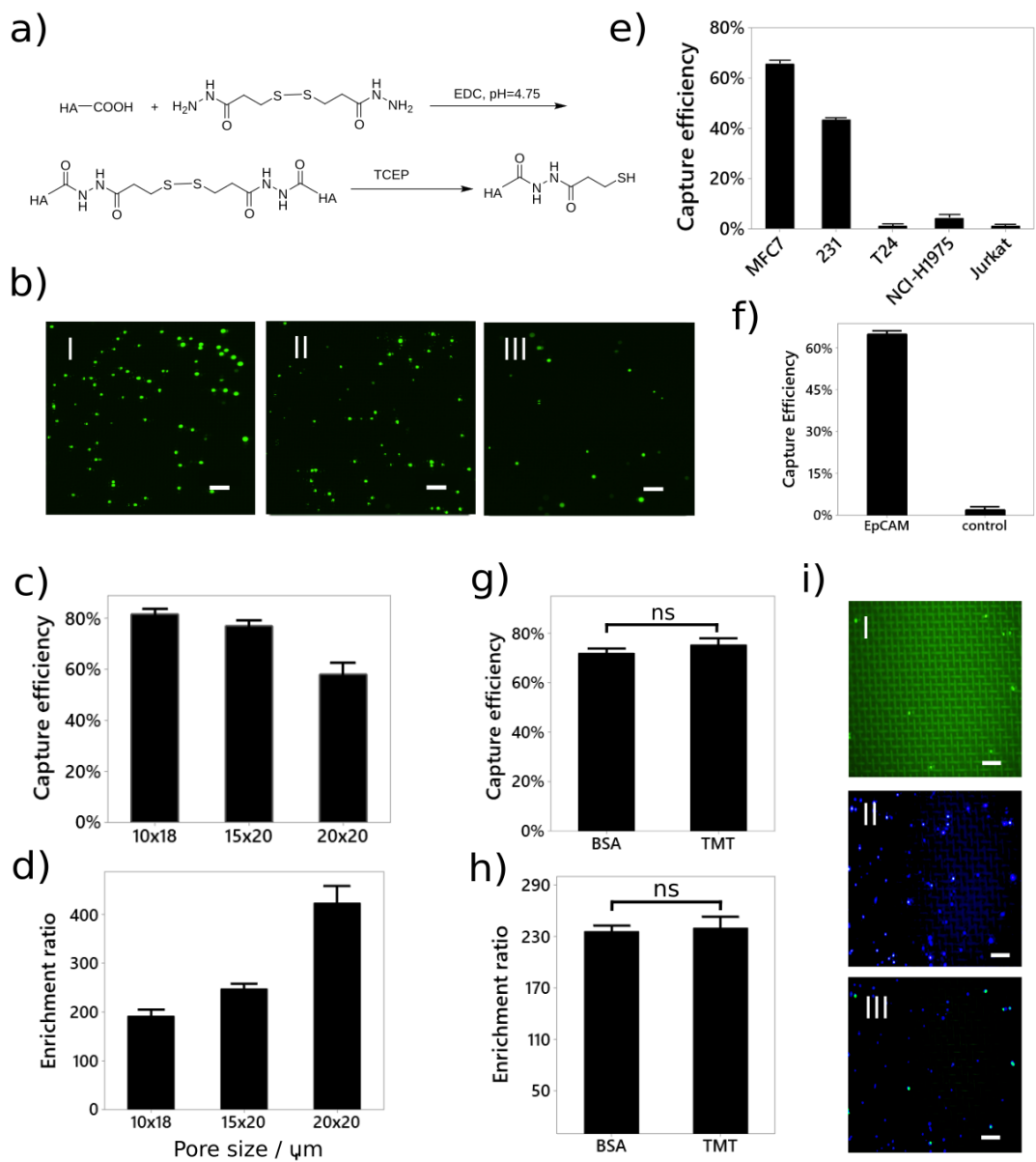

***Figure 4. Nano-functionalization strategies and performance of the device***. (***a***) *the nanobranched sulfhydryl hyaluronic acid (HA-SH) synthesis steps.* (***b***) *the micrographs of MCF-7 cells on the mesh after being captured with different concentration (approximately 150, 100 and 50 cells per mL for l, ll and III respectively) with a scale bar of 200 μm. The effect of different meshes pore sizes (10x18, 15x20 and 20x20 μm) were evaluated on cell capture efficiency (c) and enrichment ratio (****d***). (***e***) *Different cell lines exhibit capture efficiencies in line with their EpCAM expression levels (shown in Fig. S9). (****f***) *A control experiment using anti-IgG to replace anti-EpCAM antibodies shows that far fewer MCF-7 cells are captured, confirming the good immunocapture specificity.* (***g***)-(***h***)*Two blocking strategies, using bovine serum albumin (BSA) or trimethoxylsilane (TMT) have shown no significant difference in capture efficiencies or enrichment ratios.* (***i***) *shows a typical micrograph of the captured CFSE stained MCF-7 cells (l), DAPI+ cells comprised mostly of background Jurkat cells (ll) and merged image (lll), the scale bar is 200 μm.*

To further evaluate the specificity of our device, we compared the capture efficiency of MCF-7 cells with MDA-MB-231, T24, NCl-H1975 control cells (Fig. 4e). The relative capture efficiencies are in agreement with the EpCAM expression level of each cell type (Fig. S9). Figure 4f shows that far fewer MCF-7 cells are captured using anti-IgG as control on a $20 \times 20$ μm pore size micromesh, confirming the good immunocapture specificity of our approach.

Next, we quantified the enrichment of target cells, defined as the ratio of target to background cells detected (on the mesh) divided by the ratio of target to background cells in the sample[10]. For this purpose, we repeated the capture efficiency experiment with the addition of ~1 x $10^6$ Jurkat cells (EpCAM -), corresponding to a ratio of about 1:$10^4$ MCF-7: Jurkat cells. Using BSA (1%) as blocking molecules, enrichment ratios corresponding to 192, 248 and 424 were observed for $10 \times 18$ μm, $15 \times 20$ μm and 20 x 20 μm pore size meshes, respectively. These values are similar to a number of affinity-based liquid biopsies [38], [41]. The population means are significantly different at 0.05 level, using one-way ANOVA, F=28.04, confirming a reduction in non-specific interaction with decreasing total functionalized area, as expected. However, a post-hoc Tukey's test reveals that the $10 \times 18$ μm and $15 \times 20$ μm pore size mesh cannot be considered statistically different. This can be attributed to a combination of the increase in velocity reported above for the $10 \times 18$ μm pore size mesh that reduces interaction probability and the associated increased shear stress that promotes removal of non-specifically bound cells [11], [42].

We then evaluated the effect of trimethoxylsilane (50%) blocking molecules instead of BSA. Using the approach reported above, we did not observe any significant improvement in enrichment (Fig. 4h) or changes in capture efficiency (Fig. 4g) for meshes with 15 x 20 μm pore size. It is also noted that there is no significant difference in capture efficiency between the measurement with and without background cells in the same conditions (t(4) = 1.0409; p = 0.3567).

**Validation Using Clinical Samples**

Having characterized the performance of our novel device with immortalized cell lines, we validated its utility using clinical samples. In brief, we evaluated its performance in the first step of our study, then assessed its post-processing capabilities using a subset of patients (step 2) and finally compared it with clinical standards in a third step. Importantly, we note the simplicity of our set-up that can be integrated in clinical settings with minimum disruption to conventional workflows. The device only requires a single syringe pump to operate. The mesh can be easily removed from its holder, greatly simplifying pre- and post-processing procedures. In addition to relying on conventional and widely available equipment and consumables such as Petri dishes and incubators for conventional functionalization, the post-processing imaging can be directly performed on the stained meshes using epifluorescence microscopy (Fig. 1e-h).

First, we recruited 79 cancer patients and 20 healthy controls from Fudan University Shanghai Cancer Center and Changzheng hospital (ethical approval #050432-4-1911D). Demographic details of the study population are given in Table S1. To evaluate the applicability of our device and to reflect the diversity of clinical cases that could benefit from liquid biopsy, we have selected patients with 10 different cancers, including non-small-cell lung cancer and breast cancer. The volume of blood sampled was 4 mL. Meshes with 20 x 15 um pore size, 6 mm diameter, and HA-SH polymer with BSA blocking were used with the optimized flow rate of 50 μL min$^{-1}$ (as described above).

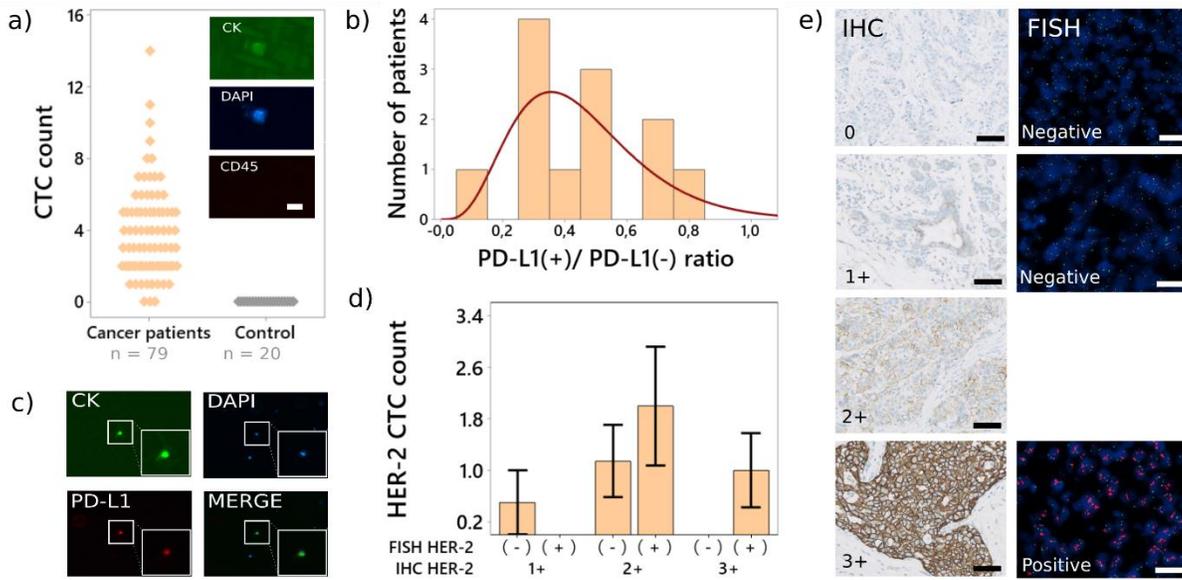

***Clinical validation of the device on two cohorts, in three separate studies.** The first study, shown in panel (**a**), with 79 cancer patients with a range of different cancer types and 20 healthy controls was used to characterize the performance of our device. CK+; DAPI+; CK- cells were identified as CTCs (inset), scale bar 20 µm. A subset of patients (n=33) from the initial cohort were selected to evaluate PD-L1 expression to identify potential responders to immune checkpoint inhibition therapy. The distribution of PD-L1+ CTCs to CTCs is shown in (**b**). PD-L1 expression was evaluated using standard secondary immunostaining (**c**), scale bar 100 µm. To evaluate our device against clinical standards, we selected 26 breast cancer patients and tested them for HER2 positivity using secondary immunostaining post-processing. We compared our results (**d**) with the histological scoring of IHC and FISH, two standard clinical assays. Representative images, with scale bar 40 µm, are shown in (**e**).*

The cells captured on the mesh (by the anti EpCAM antibody), staining for $CK^+/CD45^-/DAPI^+$, were identified as CTCs. Among the cancer patients, 96% (76/79) had at least one CTC and about 4% (3/79) had more than 10 CTCs. Using the same enumeration criteria, we tested the healthy controls and detected no CTCs (20/20). The results are summarized in Figure 5a. Using 30% as training set and 70% for classification by logistic regression, we calculated the sensitivity (96%) and specificity (100%) of our device. As confirmation, we generated the receiver operation characteristic (ROC) that yielded an area under the curve (AUC) value of 0.979 (not shown). These excellent values are due to the dual selection (immuno-capture and staining) inherent to our assay and are in line with results reported for devices using similar approaches (e.g., [11]).

In the past decade, studies on CTCs have gone beyond simple enumeration. The analysis of feature-rich CTCs, that may possess attributes of the primary tumor as well as metastasis, can provide clinically actionable information [1]–[4]. For example, the expression of PD-L1 in tumor tissues, used as a biomarker for the selection of patients eligible for immune checkpoint inhibition (ICI) therapy, is also evaluated using CTCs [43]–[45]. The upregulation of PD-L1 enables cancer cells to evade immune response by inhibiting the activation of immune cells. ICI therapies target anti PD-L1/PD1 proteins to block the inhibition of immune cells, thereby reactivating the immune system. To further demonstrate the applicability of our device in this clinical context and, in particular, to evaluate its post-processing capabilities, we selected 33 patients from the original cohort (n=79) for the evaluation of PD-L1 expression using immunofluorescence labelling (Fig. 5b,c). This step was performed after isolation of CTCs using anti-EpCAM antibodies as described earlier. On this basis, 36% of patients (12/33) had PD-L1 expressing cells. Among them, the largest proportion of patients (n=4) only had 30% of PD-L1+ CTCs as shown in Figure 5b. Based on the data from the NSCLC patients (n=21), we detected a median of 4 CTCs per 4 mL (range: 2-9 CTCs per 4 mL), among which 48% (n=10) harbored at least one PD-L1+ CTC. The total number of CTCs in 4 mL compares well with at least 2 commercial systems [46] and confirms the efficacy of our approach.

For a final validation study, we used our device to identify HER2 positive breast cancer patients based on the detection of HER2 positive CTCs. We compared our results with IHC and FISH, two standard methods approved by the U.S. Food and Drug Administration (FDA). HER2 protein overexpression and/or HER2 gene amplification, is found in about 20% of breast cancers and is associated with tumorigenesis, increased risk of metastasis, and poor prognosis [47], [48]. Importantly, these markers can be used to identify patients that could benefit from targeted therapies such as Trastuzumab, Pertuzumab and T-DM1. In this study, we selected 26 breast cancer patients from the initial cohort (n=79). They were tested for HER2 status using clinical standard procedures (IHC and FISH) performed on tissue (solid biopsy) and compared with HER2 positive CTCs captured using our device (liquid biopsy). IHC is typically used as a screening test with IHC 0 and IHC 1+ considered as negative, IHC 2+ equivocal and IHC 3+ as positive (Fig. 5d). FISH is considered more reliable, but it is more complicated and expensive, and is therefore normally used to determine the status of IHC 2+ equivocal cases [47]–[49].

**Table 1.** *Comparison between the clinical standards (IHC and FISH) and our device.*

| Subject | Patients | Patients with HER-2 positive CTC | Positive rate | Chi-square p |
|---|---|---|---|---|
| **HER2 IHC** | | | | 0.8921 |
| 1+ | 2 | 1 | 50.0% | |
| 2+ | 21 | 11 | 52.4% | |
| 3+ | 3 | 2 | 66.7% | |
| **HER2 FISH** | | | | 0.0156 |
| (-) | 16 | 5 | 31.3% | |
| (+) | 10 | 8 | 80.0% | |

The CTCs were captured using the protocol described for the PD-L1 study, with fluorescently labelled anti-ErbB2 / HER2 antibodies. The results are summarized in Figure 5c. According to the above criteria, 10 patients were identified as HER2 positive using standard approaches (IHC and/or FISH) against 14 identified based on the HER2 positive CTCs. The data shows a consistency (positive rate) of 54.17% (13/24) with the IHC and 80% (8/10, p = 0.0156) with FISH (Table 1). It is a promising result especially since FISH is considered a superior assay. We also note that our device captured HER2+ CTCs when the FISH or even IHC assay produced non-equivocal negative results (7 and 1 respectively). Similar findings were reported elsewhere[50] and may be due to intra-tumoral HER2 heterogeneity[49]. More studies, beyond the scope of this manuscript, will be necessary to evaluate the clinical relevance of HER2+ CTCs.

## Discussions

We have characterized and validated the utility of a new multiscale enrichment device that enables flow rate independent capture and processing of CTCs. We have achieved this by introducing a mesh structure that has two inherent length scales and, thus, eliminates flow rate as the main limiting factor in affinity-based liquid biopsy. The resulting device offers 1) a macroscale channel to run samples at high flow rate while maintaining low fluid velocity, 2) a microscale mesh that promotes interaction with the target cells and 3) a nano-functionalized surface that enables high capture efficiency and low non-specific interaction. Using micromeshes with 15 x 20 µm pore size and HA-SH nanobranched polymer, we have reached >76% capture efficiency, which is comparable to a number of microfluidics based liquid biopsies [37], [38]. However, unlike conventional microfluidics approaches, we have also demonstrated how our device can be scaled to allow for optimal capture efficiency for any flow rate, demonstrated here up to 12 mL h$^{-1}$ (200 µL min$^{-1}$).

Importantly, our device is easy to fabricate and assemble, its operation does not require specialist equipment and its architecture allows for simple pre- and post-processing. Hence, our approach can be seamlessly integrated into conventional laboratory workflows, including in demanding clinical environments. To demonstrate this, we validated our device using clinical samples. Using data from 79 cancer patients and 20 healthy donors as control, our device yielded 96% sensitivity and 100% specificity, which is comparable to other approaches relying on a combination of capture and staining [11], [51]. Then we used it to identify potential responders to PD-L1 ICI therapies. In comparison to commercially available liquid biopsy approaches, in the context of non-small-cell lung carcinoma, our device performed favorably[46]. Finally, we validated our approach against clinical standards in the context of HER2 positive breast cancer on a cohort of 26 breast cancer patients. In particular, we observed an 80% correspondence with FISH positive results.

Finally, we also note that our functionalization strategy is compatible with the addition of further antibodies or capture molecules for improved cell isolation efficiency [52] but also for other liquid biopsies. In addition, our nanobranched polymer is amenable to modification and may allow for the integration of cell release strategies (e.g., [26], [35]), which will enable further downstream analysis including next generation sequencing. In conclusion, our multiscale, flow- rate independent multiscale liquid biopsy approach has the potential to help drive significant advances in diagnosis, prognosis, and fundamental studies for a range of conditions.

## Materials and methods

### Device fabrication and preparation

*Fabrication.* Micromeshes of different pore sizes (10 x 18, 15 x 20 and 20 x 20 µm) were obtained from Zhongxin Hairu Ltd, China (Cat. No.: 1000 635, 800 635, 635 635, respectively). They were cleaned in 30% ethanol using ultrasound for 5 minutes, rinsed in deionized water and then dried using nitrogen ($N_2$). Meshes were gold coated (50 nm both sides) using magnetron sputtering and cut to size (e.g., 8.8 mm diameter mesh for the 6 mm diameter mesh holder) using clean surgical forceps. The mesh holders were fabricated using 3D printing or conventional machining (Fig. 1 and Fig S10).

*Preparation and cleaning of the mesh.* MilliQ water, 25% ammonium hydroxide, 30% hydrogen peroxide were mixed in a clean beaker (5:1:1 ratio, respectively) and heated to 75°C. The cut meshes were submersed for 5 minutes and washed in MilliQ water and 99% ethanol before drying with $N_2$ and then transferred to a clean petri dish for functionalization.

### Nano-functionalisation

*HA-SH nanobranched polymer synthesis.* 40 mL MES solution (Aladdin, Cat. No.: M108952, pH = 4.75, 0.1 M), was slowly added into the single-mouth flask. 200 mg sodium hyaluronate (Bloomage BioTechnology, Cat. No.: HA-TLM, molecular weight: 3.9 W) was then added into the flask and stirred (magnetic stirrer 400 rpm) until the sodium hyaluronate was fully dissolved (5-8 minutes). Then, 60 mg DTP (Frontier scientific, Cat. No.: D13817) was added into the flask and stirred thoroughly until completely dissolved. 120 mg EDC (Sinoreagent, Cat. No.: 30083834) powder was added into the solution which was then stirred at 400 rpm at room temperature for 5 h. Finally, 150 mg TCEP (Sigma, Cat. No.: C4706) was added into solution. After overnight (about 16 h) stirring, the sulfhydryl hyaluronic acid (HA-SH) was filtered (0.22 µm filter) and collected.

*Mesh functionalization.* Clean meshes (up to 18) were submerged in 3 mL HA-SH in a 5 mL centrifuge tube and orbital shook at 200 rpm for 2 h to form thiol-Au bonds between HA-SH and the mesh. After washing three time with 3 mL PBS, the meshes were submerged in 3 mL SH-PEG-COOH (Toyongbio, Cat. No.: P003002) for 1 hour to react the unbonded Au. Then the meshes were washed three times with 3 mL PBS, dried and put in a 24 well plate (one mesh per well). Activating reagents (55 µL per mesh in MES (pH = 6, 0.05 M)), comprised of 1-(3-dimethyl aminopropyl)-3-ethyl carbodiimide (EDC, Sigma, Cat. No.: 03449): 0.609 mg / mesh (35 µL) and n-hydroxysuccinimide (NHS, Sigma, Cat. No.: 56485): 0.348 mg / mesh (20 µL), were added to the surface of each HA-SH functionalized gold mesh and incubated at room temperature for 30 minutes. After incubation, the gold meshes were removed and washed three times with 500 µL PBS. The meshes were then dried and moved to new 24-well plate. The capture solution was prepared by adding 7 µL anti-EpCAM antibody (1:2000, #324202, Biolegend, CA US) to 50 µL MES solution (pH = 6, 0.05 M) and subsequent vortexing. The capture solution (57 µL) was then added onto a mesh and placed at 37°C and 5% $CO_2$ in an incubator for two hours to allow for an amide bond to be created between the anti-EpCAM antibody and HA-SH. Then, the mesh was removed and washed twice with 1 mL PBS. A blocking solution to minimize non-specific interaction (450 µL of 1% BSA solution (w/v%) – Sigma (B2064-50G)) was added to the gold mesh and returned to the incubator for one hour. After washing with PBS (Hyclone), meshes were submerged in 500 µL cryoprotectant (45% sucrose (w/v%, Sinoreagent, Cat. No.: 10021463) and 15% glycin (w/v%, Sinoreagent, Cat. No.: 62011516) in Tris-HCL (Sangon Biotech, Cat. No.: B548127-0500, 1 M, pH = 8.0)), pre-cooled to -20°C for at least one hour until it was solidified and then lyophilised. The lyophilised mesh was then sealed with desiccant and stored at -20°C ready for use.

## Cell culture and labelling

*Cell culture.* Human breast cancer (MCF-7, MDA-MB-231), urinary bladder cancer (T24), lung cancer (NCl-H1975) and monocytic (Jurkat) cells were obtained from iCell (China). All cells were cultured as recommended using phenol-red free Dulbecco's modified Eagle's medium (DMEM) (Gibco, NY, U.S.A.) supplemented with 1% l-glutamine (Life Technologies, CA, U.S.A.), 10% Fetal Bovine Serum (FBS, Gibco), and 1% penicillin/streptomycin (Corning, VA, U.S.A.), with the exception of MCF-7 which were grown in 50:50 phenol-red free DMEM:Nutrient Mixture F12 (DMEM:F12, Gibco) supplemented with 1X B27 (Gibco), 5 mg/l insulin (MBL International Corp., MA, U.S.A.), 20 µg/L basic fibroblast growth factor (bFGF, Shenandoah Inc., PA, U.S.A.), 20 µg/L epidermal growth factor (EGF, Shenandoah Inc., PA, U.S.A.), 1% penicillin/streptomycin (Corning, VA, U.S.A.), 0.5 mg/l hydrocortisone (Sigma Aldrich, MO, U.S.A.), and 2.5 mM L-glutamine (Life Technologies, U.S.A.).

*Cell labelling.* For staining, cells were trypsinised and washed twice with PBS before being stained with cell tracker (CellTrace™ CFSE Cell Proliferation Kit) following the manufacturer instructions and resuspending cells in 1 mL culture media. Green fluorescent protein (GFP) expressing MCF-7 cells were generated by lentiviral transduction with pWPI as previously described [53].

*EpCAM expression.* Flow cytometry was performed using BD Accuri C6 Flow Cytometer (BD Biosciences, U.S.A.). EpCAM mouse anti-human FITC conjugated antibody was used for epithelial marker expression (cat. # 347197, BD Bioscience, U.S.A.).

*Spiking assay.* **MCF-7 cells** grown in log phase were digested with trypsin (Life Technologies, U.S.A.), washed 2-3 times with phosphate buffer. Then, cells were accurately obtained with a cell counter and mixed into 4 mL buffer.

## Material from clinical studies

Cancer patients and control groups were recruited at Fudan University Shanghai Cancer Center and Changzheng Hospital, China (ethical approval #050432-4-1911D) after providing informed consent. Patients in this cohort may have received preoperative surgery or systematic anticancer treatment but must have been enrolled in this cohort at least 30 days in advance.

For processing in our novel devices, at least 4 mL blood of cancer patients and healthy individuals were collected and stored in EDTA tubes, blood was tested within 6 hours. Before detection of CTC, blood was processed according to the local clinical standard. Briefly, blood was diluted 1:1 in PBS (pH = 7.0) and then carefully transferred to a sterile 15 mL centrifuge tube which contained pre-warmed density gradient separation solution (4 mL, Dakewe Biotech, Shenzhen, China). This layered liquid tube was centrifuged with 700 g at room temperature for 20 minutes. The PBMC layer was pipetted into a new sterile 15 mL centrifuge tube and washed with PBS, twice (500 g, 5 minutes), and finally the PBMCs were resuspended in 300 µL PBS before use in the device.

The mesh-bound cells were fixed with 4% paraformaldehyde and washed with PBS. The fixed cells were infiltrated with 1% NP40 and blocked with 2% normal goat serum / 3% BSA. Staining to identify CTCs was performed using well established protocols using pan-Ck (Alexa Fluor488 anti-Cytokeratin (CK, pan-reactive) antibody, Biolegend (628608)), CD45 (PE anti-human CD45 Antibody, Biolegend (304008)) and DAPI (Sigma (D9542)). Secondary immunofluorescence labelled antibodies were used for the identification of PD-L1 positive cells: anti-human PD-L1 (Biolegend: 329708). Alexa Fluor 647 Anti-ErbB2 / HER2 antibody [EPR19547-12] (ab225510) was used for the identification of HER2 positive CTCs. After staining, the plate was washed with PBS and stored at 4°C until microscopic imaging.

*Tissue embedding sectioning.* The fresh patient biopsies were fixed in 4% formalin/paraformaldehyde, dehydrated in an ethanol series. After clearing in xylene, samples were infiltrated with paraffin wax. The wax block was cooled at -20°C and sliced on a microtome in 4 μm sections. For immunostaining and FISH, sections were mounted, deparaffinized and rehydrated.

*Immunohistochemistry (IHC).* Antigens were recovered in citric acid buffer (pH 6), and endogenous peroxidase activity as well as unspecific binding were blocked by transferring the sections into 3% BSA buffer. After 30 min, BSA was removed and sections were incubated with primary antibody (1:200, Alexa Fluor® 647 Anti-ErbB2 / HER2 antibody [EP1045Y] (ab281578, Abcam, British)). After overnight incubation at 4°C, sections were washed with PBS, and the secondary antibody (1:1000, Goat-Anti-Rabbit-HRP labelled, Servicebio, China) was applied at room temperature for 50 min. Again, sections were washed with PBS and stained with DAB staining kit (G1211, Servicebio, China), according to the manufacturer instructions. Nuclei in the sections were counterstained with hematoxylin. After being dehydrated and mounted, the stained tissue sections were visualized using a light microscope at x20 magnification.

*Fluorescence in situ hybridization (FISH) protocol.* The FDA approved PathVysion HER2 DNA probe kit (Abbott Molecular, IL, U.S.A.) was used according to the manufacturer protocols. Briefly, DNA on slides was denatured at $72\pm1$°C for 5 minutes and then washed and desiccated. After that, 10 μL of probe mixture was applied in a pre-warmed humidified hybridization chamber at $37\pm1$°C for 14 to 18 h. After hybridization, the sections were washed with SSC at $72\pm1$°C and desiccated in the dark. 10 μL of DAPI was applied to counterstain the sections area of the slide. Sections were observed under a fluorescence microscope. The analyses of IHC and FISH were obtained from clinicians according to clinical guidelines.

## Statistical analysis
Results were analyzed using Student's two tailed t-test, and ANOVA with equal or unequal variance in Minitab 19 (Minitab Inc., State College, PA, U.S.A.). Differences with p-values <0.05 were considered significant, and post-hoc Tukey's tests were performed after significant ANOVA differences. The logistic regression and ROC curves were obtained using the scikit-learn Python package (Python 3, on Jupyter Notebook).

## Multiphysics simulation
We used COMSOL Multiphysics (version 5.5) to conduct our simulations. To evaluate the effect of the flow rate on capture efficiency (Fig 3a), we used the "Creeping flow" and "Transport Diluted Species" modules with mesh boundary conditions (General Form Boundary PDE) to include a local Langmuir adsorption model as explained elsewhere[54]. To evaluate time-dependent discrete events (Fig3b) we selected the "Creeping flow" and "Particle tracing for Fluid flow' modules. A 'pass through' boundary condition was set on the outer perimeters of the channel, and a 'stick' condition for the mesh, so particle-wall interactions could easily be determined visually. The study on the flow velocity (Fig. 3c) was done using the Creeping flow module. Each study was repeated at least four times.

## Acknowledgements

JC would like to acknowledge funding from the Newton Fund—Institutional Links grant (ID 352360246), and Engineering and Physical Sciences Research Council grant EPSRC EP/R00403X/1. We are grateful to Hannah Bridgewater for fruitful discussions and Helena Xandri-Monje, Hadi Putra, Sunil Prasanan and Joseph Parker for contributions to initial experiments and optimization. Our gratitude goes to Professor Yin of Changzheng Hospital for sample contribution.

## Authors contribution

Conceptualisation: JY, JC. Investigation, methodology: JW, JY, RD, RL, JC. Validation: RL, YJ, RD, JC. Writing – original draft: JY, JC. Writing – review and editing: all.

## Competing interest

The authors declare the following conflicts of interests: The research was partly funded by Holosensor Medical technology Ltd and resulted in three patent applications (Y.J. and J.C. are co-inventors). J. Y. is a board member of Holosensor Medical Technology Ltd. All other authors declare they have no competing interests.

**Data and materials availability:** All processed data are available in the main text or the supplementary materials. Unprocessed data are available upon request.

## Ethics statement

The study was conducted in accordance with the Declaration of Helsinki (as revised in 2013). The study was approved by the Ethics Committee of Fudan University Shanghai Cancer Center and Changzheng hospital.

# Supplementary Materials for

## Flow Rate Independent Multiscale Liquid Biopsy for Precision Oncology

Jie Wang *et al.*

*Corresponding author. Email: jerome.charmet@he-arc.ch

**This PDF file includes:**

**Supplementary Text**

Flow rate and fluid velocity dependence in microfluidic channels

In conventional microfluidic devices, the effective cross-section A is constrained by the micron length scale in one dimension and by the microfabrication processes (e.g. aspect ratio) or practical considerations (structural stability, etc.) in the other. In other words, the flow rate Q and the fluid velocity v are coupled through the square of the length scale, as given by $v \propto \llbracket Q/L \rrbracket^2$. Therefore, any attempt to reduce the fluid velocity results in a lower interaction probability. Indeed, given a flow rate Q, the fluid velocity v can be decreased through an increase of the channel in cross-section A (as given by v=Q/A). However, this increases the length scale L (since $A \propto L^2$), which in turn decreases the interaction probability between cells and surface.

**Fig. S1.**

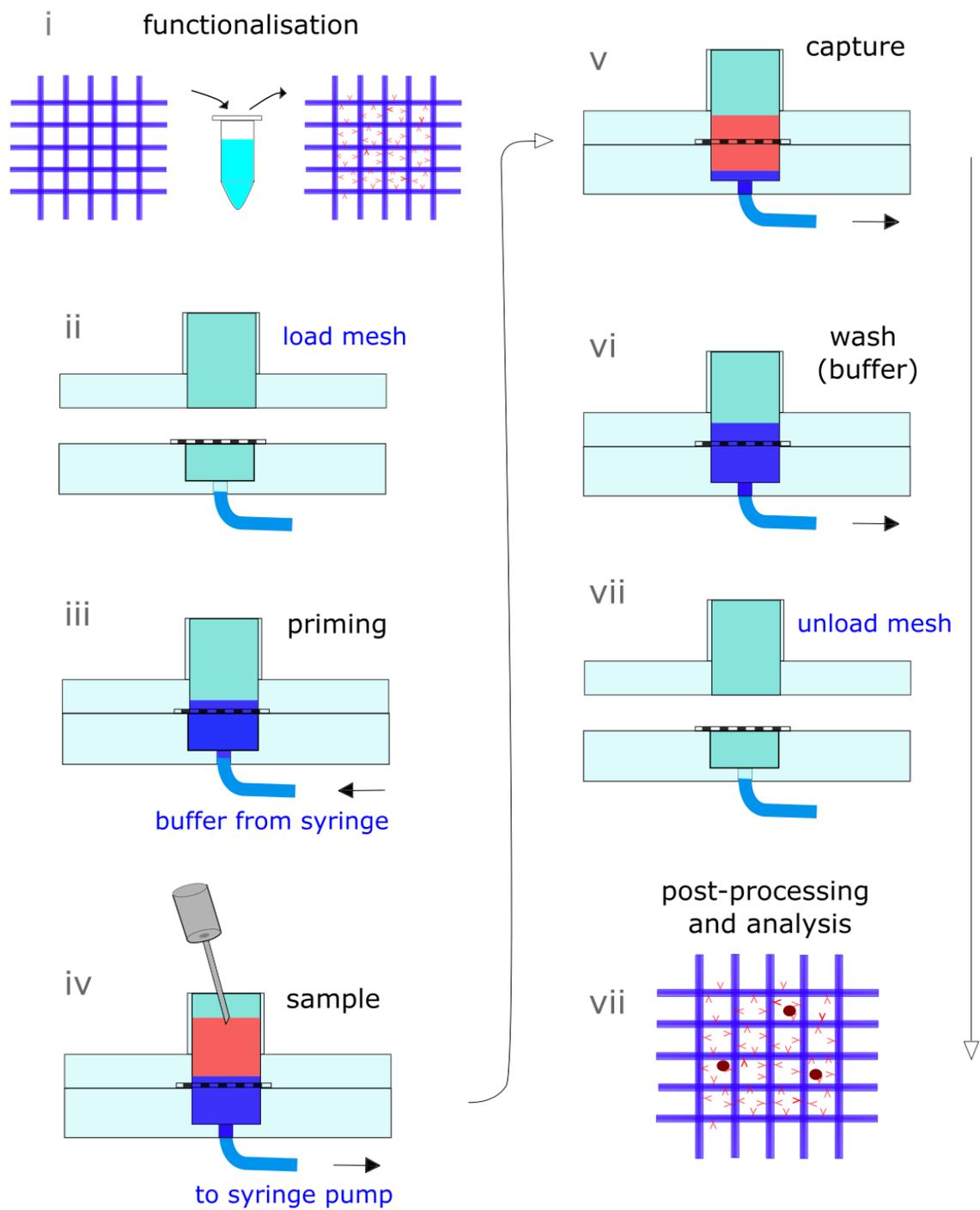

**Figure S1.** Schematic of process steps from pre-processing (functionalisation) to post-processing (e.g., fluorescence microscopy).

**Fig. S2.**

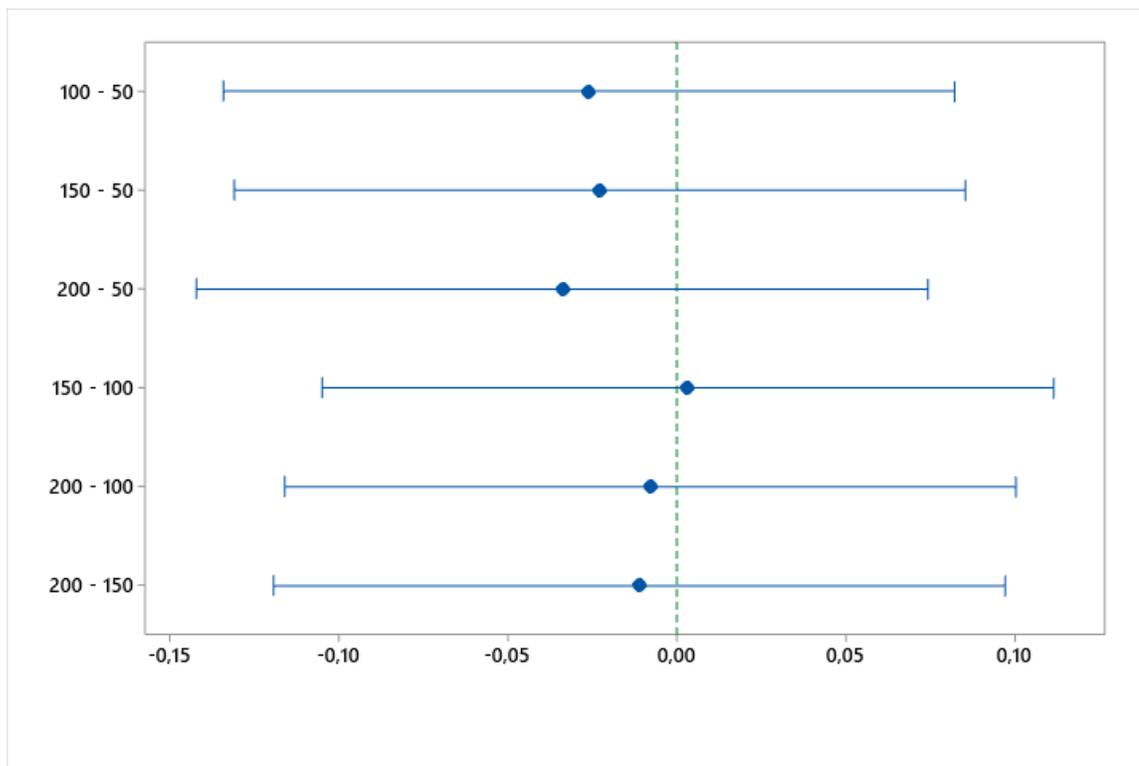

**Figure S2.** Graphs showing the post-hoc Tukey's test on flow rate independent data (Fig 2a). Each set (flow rate in µL min$^{-1}$) includes zero, which means that there is no statistical difference between any of the sets.

**Fig. S3.**

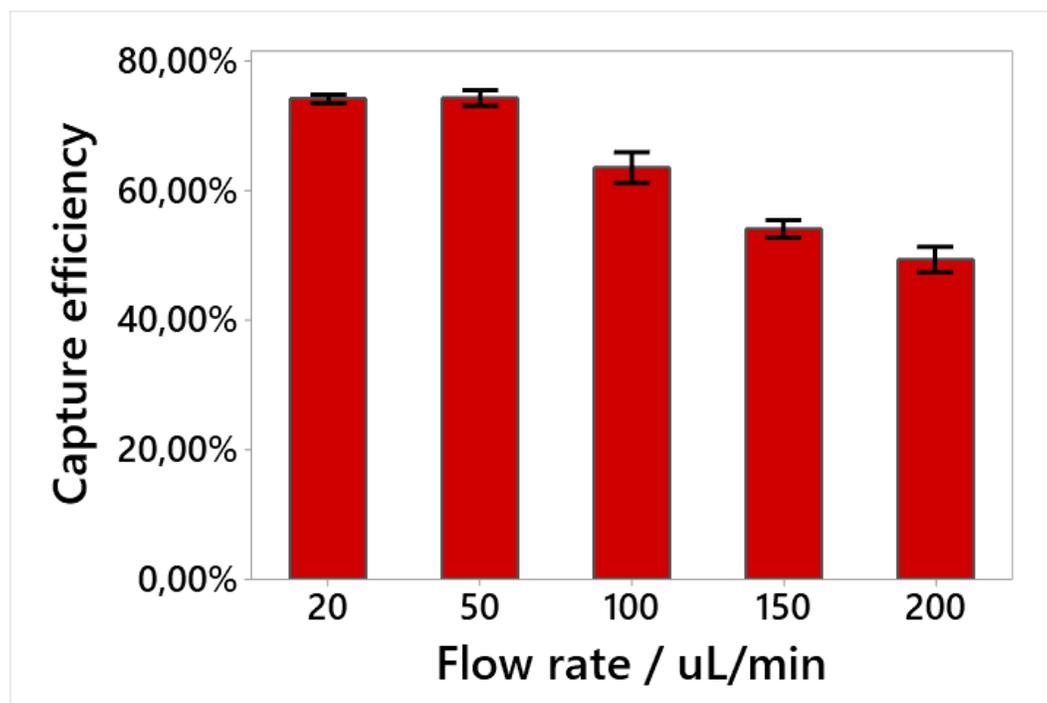

**Figure S3.** Graph showing the flow rate dependence of the device with a fixed diameter (6 mm in this case). An extra flow rate (20 µL min⁻¹) was added here to show the drop-off point at 50 µL min⁻¹, compared to Fig. 2.

**Fig. S4.**

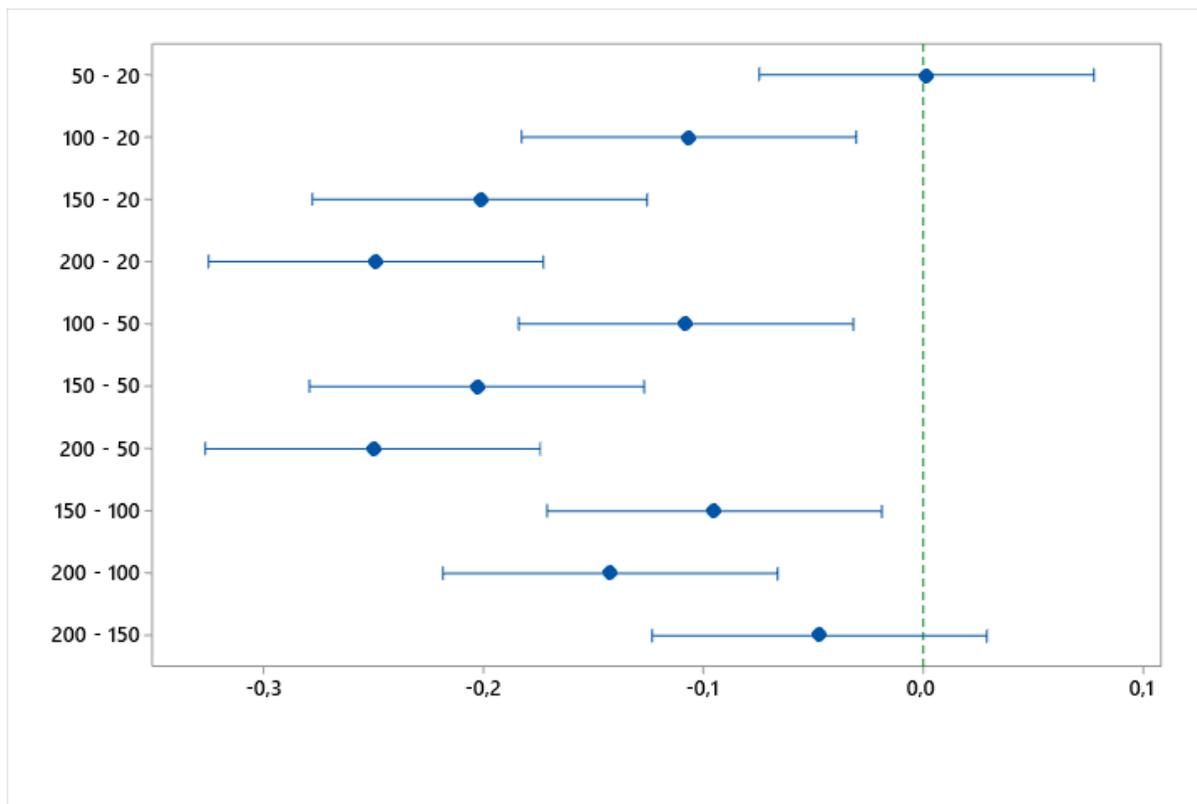

**Figure S4.** Graphs showing the post-hoc Tukey's test on flow rate dependent data (Fig S3 above). Two sets ($20 - 50$ μL min$^{-1}$ and $200 - 150$ μL min$^{-1}$) include zero, which means that there is no statistical difference between any of the sets. The rest of the sets are significantly different, confirming flow rate dependence on the capture efficiency.

**Fig. S5.**

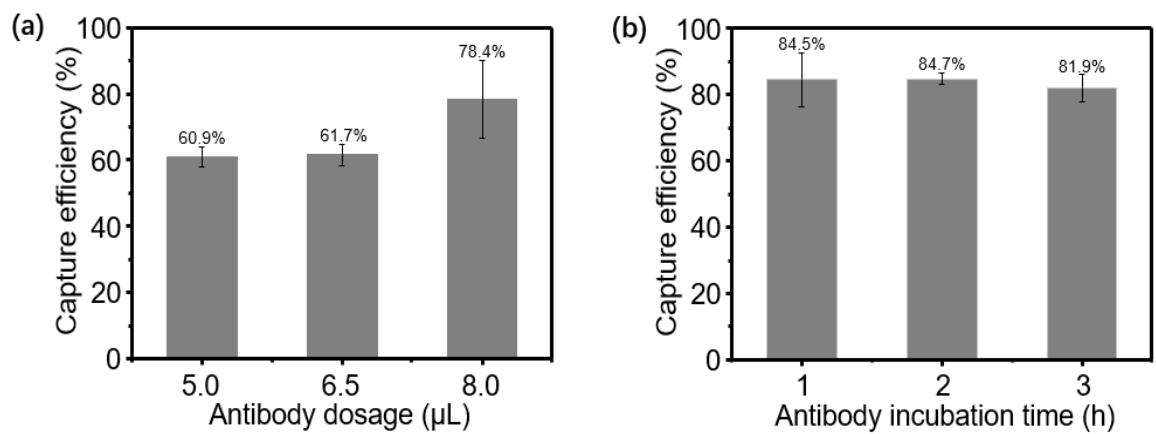

**Figure S5.** Antibody optimisation against capture efficiency (MCF-7 cells). The capture efficiency increases significantly for an 8 μL antibody dose compared 6.5 μL and 5 μL (a). This value was chosen for further analyses. Different incubation times between the polymer and anti-EpCAM antibodies were tested (b). No significant differences were observed for incubation times between 1h to 3h. The capture efficiency increases significantly for an 8 μL antibody dose compared 6.5 μL and 5 μL (a). This value was chosen for further analyses. Different incubation times between the polymer and anti-EpCAM antibodies were tested (b). No significant differences were observed for incubation times between 1h to 3h.

**Fig. S6.**

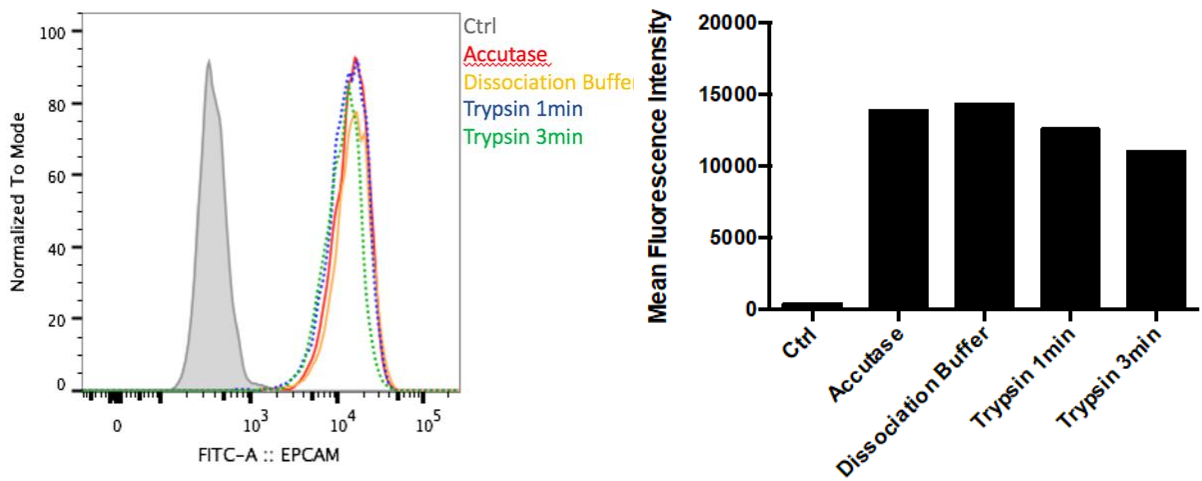

**Figure S6.** Control and optimisation of harvesting parameters and antibody selection. MCF-7 cells were stained with FITC anti-EPCAM (Clone: 9C4, Biolegend) and their EPCAM expression level was evaluated using flow cytometry. Other antibodies were tested but did not produce better results – not shown. Different harvesting reagents and parameters were tested to evaluate their effect on EpCAM expression. The effects of cell dissociation buffer in PBS, Accutase, 0.25% Trypsin (1 and 3 minutes) under standard conditions were characterised. Our results show that cell dissociation buffer in PBS or Accutase provide a gentler process, maintaining EpCAM integrity.

**Fig. S7.**

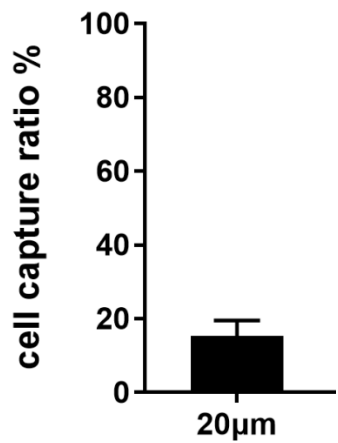

**Figure S7.** Traut reagent was used as a functionalisation strategy to tether the anti-EpCAM antibodies onto the gold coated mesh. However, the capture efficiency, presented here for a 20 x 20 μm pore size in a 6 mm diameter mesh at 50 μL min$^{-1}$, shows less than 20% capture efficiency, compared to over 60% in the same conditions with our HA-SH nanobranched polymer (Fig. 4C).

**Fig. S8.**

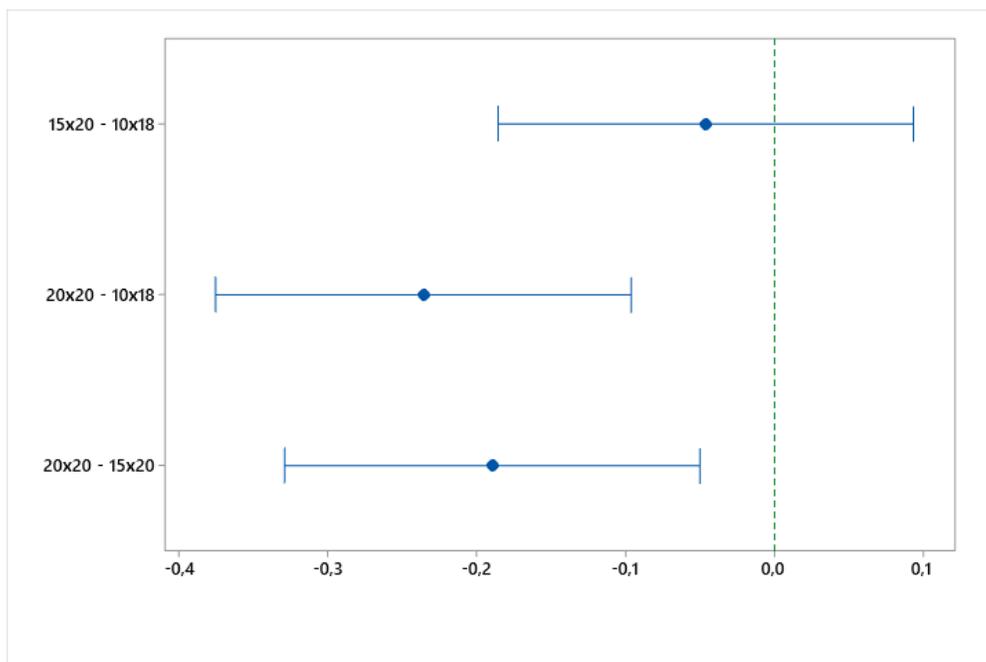

**Figure S8.** Graphs showing the post-hoc Tukey's test on the capture efficiency as a function of pore sizes. Figure 4c shows that the capture efficiency increases with decreasing pore sizes, however the graph above shows that the set (10x18 – 15x20 µm) includes zero, which means that there is no statistical difference between the two sets.

**Fig. S9.**

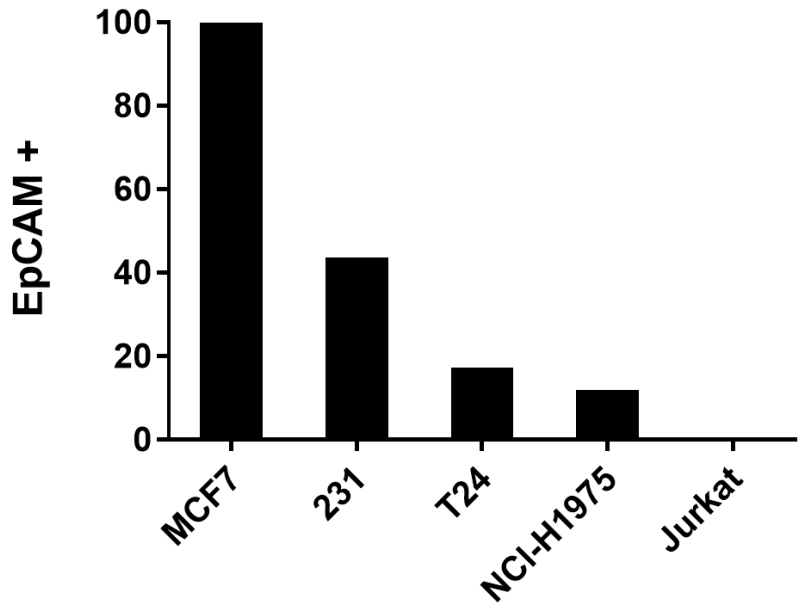

**Figure S9.** We compared the capture efficiency of MCF-7 cells with MDA-MB-231, T24, NCl-H1975 and Jurkat cells in Figure 4e. Here, we show that the relative capture efficiencies are in agreement with the EpCAM expression level of each cell type, as measured using flow cytometry. The data of each cell type is relative to the EpCAM expression level of MCF7, which is set to 100%. The results are in line with data from the gene expression atlas (Gene ID ENSG00000119888).

**Fig. S10.**

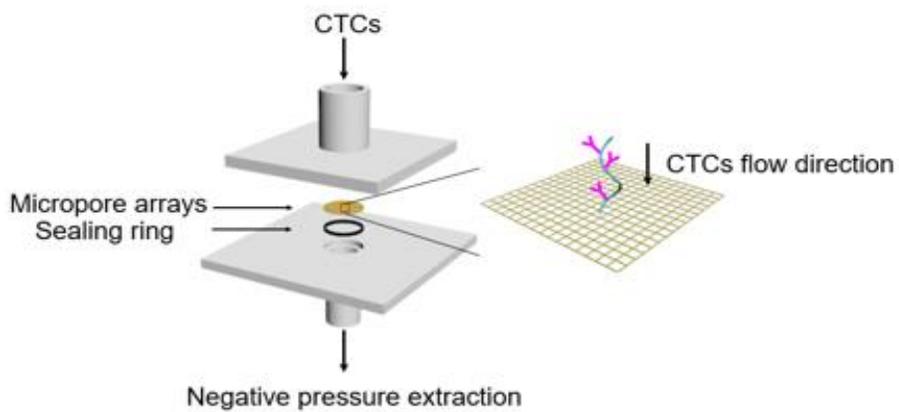

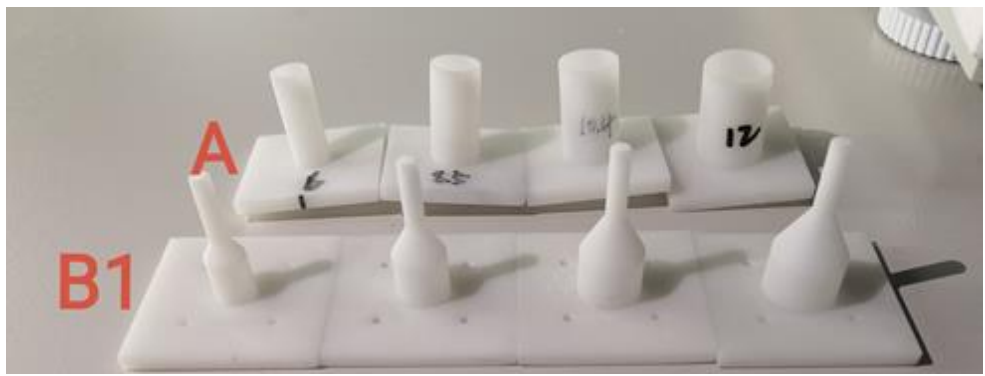

**Figure S10.** Device architecture and mesh holder. A schematic representing the assembly of the mesh in the holder (top) and different mesh holders used to evaluate the flow rate independence are shown (bottom).

**Table S1.**

| Sample | Gender | Age | Date | Diagnosis | Detection type | Number | | Rate |
| | | | | | | CTC | CTC PD-L1 | |
|---|---|---|---|---|---|---|---|---|
| 1 | F | 61 | 2019.07.29 | NSCLC | CTC+ PD-L1 | 5 | 2 | 40.0% |
| 2 | M | 53 | 2019.08.01 | NSCLC | CTC+ PD-L1 | 3 | 1 | 33.3% |
| 3 | F | 60 | 2019.08.01 | NSCLC | CTC+ PD-L1 | 2 | 0 | 0.0% |
| 4 | M | 51 | 2019.08.04 | NSCLC | CTC+ PD-L1 | 5 | 0 | 0.0% |
| 5 | M | 84 | 2019.08.25 | NSCLC | CTC+ PD-L1 | 7 | 0 | 0.0% |
| 6 | F | 49 | 2019.09.03 | NSCLC | CTC+ PD-L1 | 2 | 0 | 0.0% |
| 7 | M | 46 | 2019.09.05 | NSCLC | CTC+ PD-L1 | 6 | 2 | 33.3% |
| 8 | F | 58 | 2019.11.24 | NSCLC | CTC+ PD-L1 | 6 | 0 | 0.0% |
| 9 | M | | 2019.11.25 | NSCLC | CTC+ PD-L1 | 4 | 0 | 0.0% |
| 10 | M | | 2019.11.28 | NSCLC | CTC+ PD-L1 | 4 | 0 | 0.0% |
| 11 | F | 47 | 2019.12.09 | NSCLC | CTC+ PD-L1 | 2 | 1 | 50.0% |
| 12 | M | 68 | 2019.12.26 | NSCLC | CTC+ PD-L1 | 4 | 1 | 25.0% |
| 13 | M | | 2019.12.27 | NSCLC | CTC+ PD-L1 | 2 | 1 | 50.0% |
| 14 | M | | 2019.12.28 | NSCLC | CTC+ PD-L1 | 2 | 1 | 50.0% |
| 15 | F | | 2019.12.28 | NSCLC | CTC+ PD-L1 | 9 | 1 | 11.1% |
| 16 | F | | 2020.01.10 | NSCLC | CTC+ PD-L1 | 3 | 1 | 33.3% |
| 17 | M | | 2020.01.10 | NSCLC | CTC+ PD-L1 | 2 | 0 | 0.0% |
| 18 | M | | 2020.01.10 | NSCLC | CTC+ PD-L1 | 3 | 0 | 0.0% |
| 19 | M | | 2020.03.16 | NSCLC | CTC+ PD-L1 | 6 | 0 | 0.0% |
| 20 | | | 2020.03.23 | NSCLC | CTC+ PD-L1 | 4 | 0 | 0.0% |
| 21 | M | 56 | 2020.04.17 | colorectal cancer | CTC+ PD-L1 | 3 | 0 | 0.0% |
| 22 | F | 62 | 2020.04.17 | colorectal cancer | CTC+ PD-L1 | 3 | 2 | 66.7% |
| 23 | F | 64 | 2020.05.09 | colorectal cancer | CTC+ PD-L1 | 4 | 0 | 0.0% |
| 24 | F | | 2020.05.26 | NSCLC | CTC+ PD-L1 | 5 | 4 | 80.0% |
| 25 | M | 57 | 2020.06.05 | SCLC | CTC+ PD-L1 | 3 | 2 | 66.7% |
| 26 | | | 2020.06.05 | liver cancer | CTC+ PD-L1 | 6 | 0 | 0.0% |
| 27 | M | 51 | 2020.06.14 | Pulmonary nodules | CTC+ PD-L1 | 3 | 0 | 0.0% |
| 28 | | | 2020.06.05 | liver cancer | CTC+ PD-L1 | 2 | 0 | 0.00% |
| 29 | | | 2020.06.05 | liver cancer | CTC+ PD-L1 | 6 | 0 | 0.00% |
| 30 | | | 2020.06.30 | liver cancer | CTC+ PD-L1 | 2 | 0 | 0.00% |
| 31 | F | 76 | 2022.01.14 | liver cancer | CTC+ PD-L1 | 1 | 0 | 0.00% |
| 32 | M | 36 | 2022.02.15 | liver cancer | CTC+ PD-L1 | 1 | 0 | 0.00% |
| 33 | F | 74 | 2022.02.10 | pancreatic carcinoma | CTC+ PD-L1 | 0 | 0 | 0.00% |

**Table S1 (a):** Details of patients enrolled in the first and second studies and raw results for CTCs and PD-L1+ CTCs counts.

**Table S2.**

| Sample | Gender | Age | Date | Diagnosis | Detection type | CTC number |
|--------|--------|-----|------|-----------|----------------|------------|
| 1 | M | | 2019.08.23 | NSCLC | CTC | 6 |
| 2 | M | | 2019.08.23 | NSCLC | CTC | 4 |
| 3 | M | | 2019.08.27 | NSCLC | CTC | 1 |
| 4 | M | | 2019.08.27 | NSCLC | CTC | 3 |
| 5 | M | | 2019.08.27 | NSCLC | CTC | 2 |
| 6 | M | | 2019.08.27 | NSCLC | CTC | 1 |
| 7 | M | | 2019.08.27 | NSCLC | CTC | 5 |
| 8 | M | | 2019.08.30 | NSCLC | CTC | 4 |
| 9 | F | | 2019.08.30 | NSCLC | CTC | 4 |
| 10 | M | | 2019.08.30 | NSCLC | CTC | 2 |
| 11 | M | | 2019.08.30 | NSCLC | CTC | 4 |
| 12 | F | | 2019.09.13 | NSCLC | CTC | 5 |
| 13 | M | | 2019.10.26 | NSCLC | CTC | 2 |
| 14 | M | 42 | 2019.12.24 | nasopharynx cancer | CTC | 14 |
| 15 | F | | 2020.03.14 | cancer of biliary duct | CTC | 3 |
| 16 | | | 2020.06.30 | liver cancer | CTC | 2 |
| 17 | F | 40 | 2019.11.21 | breast cancer | CTC | 4 |
| 18 | F | 51 | 2020.01.02 | breast cancer | CTC | 5 |
| 19 | F | 68 | 2020.03.12 | breast cancer | CTC | 2 |
| 20 | F | 66 | 2020.03.12 | breast cancer | CTC | 2 |

| Number | Sampling time | Age | Gender | CTC |
|--------|---------------|-----|--------|-----|
| 1 | 2021.12.04 | 29 | Male | 0 |
| 2 | 2021.12.04 | 28 | Male | 0 |
| 3 | 2021.12.04 | 31 | Female | 0 |
| 4 | 2021.11.25 | 24 | Male | 0 |
| 5 | 2021.11.25 | 29 | Male | 0 |
| 6 | 2021.11.25 | 35 | Male | 0 |
| 7 | 2021.11.26 | 24 | Female | 0 |
| 8 | 2021.11.26 | 28 | Female | 0 |
| 9 | 2021.11.26 | 29 | Male | 0 |
| 10 | 2021.12.28 | 31 | Female | 0 |
| 11 | 2021.12.28 | 31 | Male | 0 |
| 12 | 2021.12.28 | 30 | Male | 0 |
| 13 | 2021.12.28 | 33 | Male | 0 |
| 14 | 2021.12.28 | 32 | Male | 0 |
| 15 | 2021.12.19 | 28 | Male | 0 |
| 16 | 2021.12.19 | 26 | Male | 0 |
| 17 | 2021.12.19 | 29 | Female | 0 |
| 18 | 2021.12.19 | 24 | Female | 0 |
| 19 | 2021.12.20 | 27 | Male | 0 |
| 20 | 2021.12.20 | 27 | Male | 0 |

**Table S1 (b):** Details of patients enrolled in the first study and raw results for CTCs counts only (top) and healthy volunteers (bottom).

**Table S2.**

| Time | NO. | Age | Diagnosis | | | | Tissue grade | T | N | M | TNM | IHC HER2 | FISH HER-2 | ER | PR | Size | CTC | HER-2 CTC |
|------|-----|-----|-----------|---|---|---|---|---|---|---|-----|----------|-----------|-----|-----|------|-----|-----------|
| 2020.09.24 | 1 | 74 | Breast cancer | Malignant | Invasive | Ductal carcinoma | 3 | 2 | 1 | 0 | ⅡB | 2+ | (-) | 80%，+ | 60%，+ | 1.3*1.3*1cm | 2 | 0 |
| 2020.11.06 | 2 | 41 | Breast cancer | Malignant | Invasive | Ductal carcinoma | 2 | 0 | 0 | 0 | 0 | 2+ | (-) | 95%，+ | 60%，+ | / | 1 | 0 |
| 2020.11.06 | 3 | 77 | Breast cancer | Malignant | Invasive | Ductal carcinoma | 3 | 2 | 0 | 0 | ⅡA | 2+ | (-) | 80%，+ | 80%，+ | 2*1.8*1.5cm | 1 | 0 |
| 2020.11.13 | 4 | 48 | Breast cancer | Malignant | Invasive | Ductal carcinoma | 3 | 2 | 0 | 0 | ⅡA | 3+ | (+) | - | - | 4*3.5*2cm | 5 | 2 |
| 2020.11.13 | 5 | 60 | Breast cancer | Malignant | Invasive | Ductal carcinoma | 2 | 3 | 1 | 0 | ⅢA | 2+ | (+) | - | - | 5×4×2cm | 11 | 6 |
| 2020.11.13 | 6 | 71 | Breast cancer | Malignant | Invasive | Apocrine sweat gland carcinoma | 2 | 1 | 0 | 0 | Ⅰ | 2+ | (-) | - | - | 1.2*0.9*0.8cm | 8 | 6 |
| 2020.11.17 | 7 | 59 | Breast cancer | Malignant | / | / | / | / | / | / | / | 2+ | (-) | 90%，+ | 80%，+ | / | 5 | 0 |
| 2020.12.02 | 8 | 33 | Breast cancer | Malignant | Invasive | Ductal carcinoma | 2 | / | / | / | / | 2+ | (-) | 80%，+ | 90%，+ | 2.5×1.5×1.5cm | 10 | 6 |
| 2020.12.02 | 9 | 63 | Breast cancer | Malignant | Invasive | Ductal carcinoma | 2 | 1 | 0 | 0 | Ⅰ | 2+ | (+) | 80%，+ | 1% | 2×1×1cm | 8 | 5 |
| 2020.12.03 | 10 | 41 | Breast cancer | Malignant | Invasive | Ductal carcinoma | 2 | 3 | 0 | 0 | ⅡB | 2+ | (-) | 80%，+ | 80%，+ | 6×3×2cm | 5 | 0 |
| 2020.12.03 | 11 | 42 | Breast cancer | Malignant | Invasive | Ductal carcinoma | 2 | / | / | / | / | 2+ | (-) | 70%，+ | 70%，+ | 1.6×1.1×1.3cm | 5 | 1 |

| Time | NO. | Age | Diagnosis | | | | Tissue grade | T | N | M | TNM | IHC HER2 | FISH HER-2 | ER | PR | Size | CTC | HER-2 CTC |
|---|---|---|---|---|---|---|---|---|---|---|---|---|---|---|---|---|---|---|
| 2020.12.07 | 12 | 56 | Breast cancer | Malignant | Invasive | Ductal carcinoma | 3 | / | / | / | / | 2+ | (-) | - | - | 2×1.5×1 cm | 2 | 1 |
| 2020.12.08 | 13 | 31 | Breast cancer | Malignant | Invasive | Ductal carcinoma | 3 | / | / | / | / | 2+ | (+) | 90%，+ | 80%，+ | / | 7 | 1 |
| 2020.12.09 | 14 | 27 | Breast cancer | Malignant | / | / | / | 2 | 2 | 1 | Ⅳ | 2+ | (+) | - | - | / | 0 | 0 |
| 2020.12.10 | 15 | 49 | Breast cancer | Malignant | / | / | / | / | / | / | / | 3+ | (+) | 70%，+ | 2%， | 25×25×15 mm | 3 | 0 |
| 2020.12.11 | 16 | 36 | Breast cancer | Malignant | Invasive | Ductal carcinoma | 3 | 1 | 1 | 0 | ⅡA | 2+ | (-) | - | - | 1.1×1×0.9cm | 1 | 1 |
| 2020.12.11 | 17 | 48 | Breast cancer | Malignant | Invasive | Ductal carcinoma | 2 | 1 | 0 | 0 | Ⅰ | 2+ | (-) | 80%，+ | 80%，+ | 1.5×1×1 cm | 5 | 0 |
| 2020.12.14 | 18 | 41 | Breast cancer | Malignant | Invasive | Ductal carcinoma | 2--3 | 2 | 1 | 0 | ⅡB | 2+ | (+) | 80%，+ | 80%，+ | 2.5×1.5×1.3 cm | 5 | 0 |
| 2020.12.14 | 19 | 68 | Breast cancer | Malignant | Invasive | Ductal carcinoma | 3 | / | / | / | / | 2+ | (+) | - | - | 1.4×1.2×1cm | 5 | 1 |
| 2020.12.15 | 20 | 56 | Breast cancer | Malignant | Invasive | Ductal carcinoma | 2 | 1 | 0 | 0 | Ⅰ | 2+ | (-) | 60%，+ | 1%，+ | 2×1.2×1.1cm | 7 | 1 |
| 2020.12.15 | 21 | 65 | Breast cancer | Malignant | Invasive | Ductal carcinoma | 3 | 2 | 1 | 0 | ⅡB | 2+ | (+) | 90%，+ | 2% | 2×2×2 cm | 2 | 1 |
| 2020.12.16 | 22 | 36 | Breast cancer | Malignant | Invasive | / | / | / | / | / | / | 1+ | (-) | 10% | - | / | 7 | 0 |
| 2020.12.21 | 23 | 47 | Breast cancer | Malignant | Invasive | Ductal carcinoma | 2 | x | 0 | 0 | | 2+ | (-) | 80%，+ | 80%，+ | / | 0 | 0 |

| Time | NO. | Age | Diagnosis | | | | Tissue grade | T | N | M | TNM | IHC HER2 | FISH HER-2 | ER | PR | Size | CTC | HER-2 CTC |
|---|---|---|---|---|---|---|---|---|---|---|---|---|---|---|---|---|---|---|
| 2020.12.22 | 24 | 49 | Breast cancer | Malignant | Invasive | Ductal carcinoma | 3 | / | / | / | / | 3+ | (+) | - | - | 2*2*1cm | 7 | 1 |
| 2020.12.24 | 25 | 50 | Breast cancer | Malignant | Invasive | / | 2 | / | / | / | / | 2+ | (-) | 90%，+ | 90%，+ | 2×2×1cm | 3 | 0 |
| 2020.12.28 | 26 | 43 | Breast cancer | Malignant | Invasive | Lobular carcinoma | 2 | 1 | 1 | 0 | ⅡA | 1+ | (-) | 80%，+ | 80%，+ | 2.5×2×1.5cm | 1 | 1 |